\documentstyle[prb,aps,multicol,epsfig]{revtex}

\begin{document}
\draft

\title{Excitons in one-dimensional Mott insulators}

\author{F.H.L.~Essler$^1$, F.~Gebhard$^2$ and E.~Jeckelmann$^2$}
\address{$^1$ Department of Physics, Warwick University,
Coventry, CV4 7AL, UK\\
$^2$ Fachbereich Physik, Philipps-Universit\"at Marburg, D-35032 
Marburg, Germany}

\maketitle

\begin{abstract}
We employ dynamical density-matrix renormalization group (DDMRG) and 
field-theory methods to determine the frequency-dependent optical
conductivity in one-dimensional extended, half-filled Hubbard models.
The field-theory approach is applicable to the regime of ``small'' Mott
gaps which is the most difficult to access by DDMRG.
For very large Mott gaps the DDMRG recovers analytical results obtained
previously by means of strong-coupling techniques. 
We focus on exciton formation at energies below the onset of the
absorption continuum. As a consequence of spin-charge separation, these
Mott--Hubbard excitons are bound states of spinless, charged excitations
(``holon--antiholon'' pairs). We also determine exciton
binding energies and sizes. 
In contrast to simple band
insulators, we observe that excitons exist in the Mott-insulating
phase only for a sufficiently strong intersite Coulomb
repulsion. Furthermore, our results show that the exciton binding
energy and size are not related in a simple way to the strength of the 
Coulomb
interaction. 
\end{abstract}


\begin{multicols}{2}
\narrowtext
\section{Introduction}
\label{Sec:Introduction}

Excitons in conventional band insulators are well-described by
Wannier theory~\cite{Koch}. A Wannier exciton is a charge neutral
optical excitation made of an electron in the conduction band and a 
hole in
the valence band, bound together by the Coulomb attraction between
them. In inorganic semiconductors like GaAs the typical binding
energy, as defined by the energy difference between the exciton and
the band edge of the particle-hole continuum, is several meV. 
This should be compared to the band gap itself, which is of the order
of one~eV.
Concomitantly the typical size of a Wannier exciton is of the order of
100~\AA, which is almost two orders of magnitude larger than the lattice
spacing. We note that although the total spin of an optically excited
exciton is necessarily zero, it is composed of two quasiparticles
carrying spin-1/2.

In quasi one-dimensional materials like conjugated 
polymers~\cite{Farges} 
the situation is quite different.
Here the electron-electron
interaction accounts, to a substantial degree, for the optical gap
itself~\cite{Dionys} as well as for the formation of excitons.
The exciton binding energy in, e.g., polydiacetylenes is found to be of
the order of 0.5~eV~\cite{Weiser1,Weiser2} 
and is thus comparable to the optical gap of 
2.4~eV in 3BCMU-polydiacetylene chains diluted in their monomer 
matrix~\cite{Weiser1,Weiser2}.  
The exciton size was estimated~\cite{Weiser2} to be 
12~\AA\ 
and is thus comparable to the
length of the unit cell of~5~\AA. 
These facts suggest that electron-electron
interactions will play an important role in any theoretical
description of excitons in these materials. 

Realistic models for conjugated polymers must account for the
effects of both electron-electron and electron-phonon
interactions. The interplay between these makes the reliable
calculation of the optical conductivity a very demanding task.
As a first step it is therefore natural to investigate the effects of
the two mechanisms separately~\cite{Sari}. The optical conductivity
for models with
only electron-lattice coupling such as the celebrated
Su-Schrieffer-Heeger model has been widely analyzed in the
literature~\cite{SSH} and is well established. As an extension
of this approach, electron-electron interactions have been taken into 
account
perturbatively~\cite{AbeSchreiber}.
In contrast 
there are comparatively few reliable results for
the optical spectrum in models that take account of the sizable 
electron-electron
interaction~\cite{mazumdar,bursill,shuai,Rama1}. 
The interaction drives these
systems into a Mott-insulating ground state~\cite{Mott}. The scarcity
of results is due to the difficulties associated with the calculation
of excited state properties in one-dimensional Mott insulators.
Therefore, it is of considerable interest
to develop reliable methods for the investigation
of optical excitations in correlated electron
systems, and to determine the optical conductivity of 
one-dimensional Mott insulators.

In this paper we focus on the calculation of the optical conductivity
in models with electron-electron interactions only. We study an
extended Hubbard model with nearest and next-nearest neighbor
density-density interactions; the model is further specified in 
Sec.~\ref{Sec:model}.

We employ two recently developed numerical
and analytical techniques to determine the real part of the
optical conductivity over the full
frequency range and analyze exciton properties without suffering from
finite-size limitations; for a first application to the Hubbard model,
see Ref.~\onlinecite{JGE}.   
In Sec.~\ref{Sec:DMRG} we first test the dynamical 
density-matrix renormalization group (DDMRG) by applying it to the 
limit of
large Mott gaps where analytical results are
available~\cite{Marburger}. We obtain excellent agreement between
numerical and analytical results, and confirm the 
clear and simple physical picture of an exciton as a bound state of
a double occupancy and an empty lattice site in a background of 
singly occupied sites. We then move on to the generic case of
intermediate Mott gaps and find qualitatively the same
physical behavior.

In the regime of small Mott gaps, finite-size effects and finite
resolution of the DDMRG start to hamper the numerical analysis.
Therefore, in Sec.~\ref{Sec:FT}, we carry out a weak-coupling
field-theory analysis of the problem. Using the form-factor bootstrap
approach we determine the optical conductivity in the field-theory
limit. Here, the spin sector does not couple to the current operator
so that it is sufficient to analyze the charge sector only.
The exciton is then described as a bound state of a holon 
and an antiholon, which are the elementary charge excitations in the
theory. The resulting picture for small Mott gaps remains very
similar to the cases of intermediate to large gaps. 
We even find quantitative agreement
between field theory and DDMRG results for intermediate Mott gaps
where the applicability of field theory is not a priori expected.

In Sec.~\ref{Sec:excitonproperties} we discuss two fundamental
properties of Mott--Hubbard excitons, 
their binding energy and size,
in greater detail. 
In contrast to Wannier excitons in band insulators,
Mott--Hubbard excitons exist only when the intersite Coulomb
repulsion exceeds a certain threshold.  
In general, the exciton binding energy is not related in a simple way 
to the strength of the Coulomb interaction. 
We analyze a new correlation function which allows to
define the size of an exciton in correlated electron systems.
Our analysis provides a comprehensive 
picture of excitons in one-dimensional
Mott insulators.

In our conclusions (Sec.~\ref{Sec:conclusions}) we address implications
of our results for the theory of $\pi$-conjugated polymers.

\section{Model Hamiltonian}
\label{Sec:model}

In this work we study the one-dimensional extended Hubbard 
model~\cite{Hubbard},
\begin{eqnarray}
\hat{H} 
&=& -t \sum_{l;\sigma} \left( \hat{c}_{l,\sigma}^+\hat{c}_{l+1,\sigma} 
+ \hat{c}_{l+1,\sigma}^+\hat{c}_{l,\sigma} \right) \nonumber \\
&&+ U \sum_{l} \left(\hat{n}_{l,\uparrow}-\frac{1}{2}\right)
\left(\hat{n}_{l,\downarrow}-\frac{1}{2}\right)  \nonumber \\
&& + V_1 \sum_{l}(\hat{n}_l-1)(\hat{n}_{l+1}-1)\label{Hamiltonian}\\
&& + V_2\sum_{l}(\hat{n}_l-1)(\hat{n}_{l+2}-1) \nonumber \, .
\end{eqnarray}
This Hamiltonian describes electrons with spin 
$\sigma=\uparrow,\downarrow$ 
which can hop between neighboring sites. Here $\hat{c}^+_{l,\sigma}$,
$\hat{c}_{l,\sigma}$ are creation and annihilation operators for
electrons with spin $\sigma$ at site $l$, $\hat{n}_{l,\sigma}=
\hat{c}^+_{l,\sigma}\hat{c}_{l,\sigma}$ are the corresponding number
operators, and $\hat{n}_l=\hat{n}_{l,\uparrow}+\hat{n}_{l,\downarrow}$.

Since we are interested
in the Mott insulating phase, we exclusively consider a half-filled band
where the number of electrons~$N$ equals the
number of lattice sites~$L$. The lattice spacing is set to unity,
$a_0\equiv 1$.
Note that we have chosen the chemical potential in such a way 
that the Hamiltonian explicitly exhibits a particle-hole symmetry.
This Hamiltonian has two other discrete symmetries which are useful
for optical excitation calculations: a spin-flip symmetry
and a spatial-reflection symmetry (through
the lattice center). 
Therefore, each eigenstate of (\ref{Hamiltonian}) has
a well-defined parity under charge conjugation
($P_{\rm c} = \pm 1$) and spin flip ($P_{\rm s} = \pm 1$), and
belongs to one of the two irreducible
representations, $A_g$ or $B_u$, of a   
one-dimensional lattice reflection 
symmetry group.  

The kinetic energy is diagonal in momentum space and gives rise to a 
cosine band,
$\epsilon(k)=-2t\cos(k)$ of width~$W=4t$.
The Coulomb repulsion is
mimicked by a repulsive, local Hubbard interaction~$U$, and nearest and
next-nearest neighbor repulsions~$V_1$, $V_2$. We restrict ourselves
to the physically relevant case $U >  V_1 > V_2 \geq 0$.
In this work, we are not interested in issues like a complete 
classification
of the phase diagram of the model~(\ref{Hamiltonian});
instead, we constrain our analysis to the consideration of
several different points in the Mott insulating phase.
A more systematic investigation of the extended Hubbard model with
$V_2=0$ will be published elsewhere~\cite{EricUV}. There
it is shown that without the next-nearest neighbor interaction,
it is not possible to have simultaneously a small Mott gap and form
a Mott--Hubbard exciton.

Linear optical absorption is one of the most
commonly used probes in experimental studies 
of the dynamical properties of a material.
The optical absorption is proportional 
to the real part of the optical conductivity,
which is related to the imaginary part of the current-current 
correlation
function by
\begin{mathletters}%
\label{decomp}
\begin{eqnarray}
\sigma_1(\omega>0) &=&\frac{{\rm Im}\{\chi(\omega>0)\}}{\omega} \; ,
\label{decompA}\\
\chi(\omega>0) 
&=& - \frac{1}{L} \langle 0|\hat{\jmath} 
\frac{1}{E_0-\hat{H}+\hbar\omega+i\eta}\hat{\jmath} |0\rangle 
\label{decompB}\\
&=& - \frac{1}{L} \sum_n 
\frac{ |\langle 0 | \hat{\jmath} | n\rangle|^2}{\hbar\omega -(E_n-E_0) 
+i\eta}
 \; .
\label{decompC}
\end{eqnarray}\end{mathletters}%
Here, $|0\rangle$ is the ground state, $|n\rangle$ are excited states,
and $E_0$, $E_n$ are their respective energies. 
Although $\eta=0^+$ is infinitesimal,
we may introduce a finite value to broaden our resonances at 
$\hbar\omega=E_n-E_0$.
In momentum space, the current operator reads
\begin{equation}
\hat{\jmath}=-\frac{2et}{\hbar}  \sum_{k;\sigma} \sin(k)
\hat{c}_{k,\sigma}^+\hat{c}_{k,\sigma}\; .
\label{currentop}
\end{equation} 
We note that the current operator is invariant under the 
spin-flip transformation but antisymmetric under 
charge conjugation and spatial reflection. 
Therefore, if the ground state $|0\rangle$ belongs to
the symmetry subspace ($A_g,P_{\rm c},P_{\rm s}$), only excited states
$|n\rangle$ belonging to the symmetry subspace
($B_u,-P_{\rm c},P_{\rm s}$) contribute to the optical conductivity.
According to selection rules, the matrix element 
$\langle 0 | \hat{\jmath} | n\rangle$ vanishes if $|n\rangle$
belongs to another symmetry subspace.
We set $\hbar = 1$ throughout, and for the presentation of 
our results we use $e = t \equiv  1$ in our figures.

\section{Density-matrix renormalization group}
\label{Sec:DMRG} 

Recently, the density-matrix renormalization group method 
\cite{Steve,DMRGbook} (DMRG) has been extended to the calculation
of {\sl dynamical\/} correlation functions
\cite{Rama1,JGE,KuehnerWhite}. This numerical technique allows us to
obtain $\sigma_1(\omega)$ for all interaction strengths as long as the
gap is not exponentially small. 
A complete exposition of our DDMRG method  
will be published elsewhere~\cite{Eric2}.

DDMRG allows us to calculate dynamical
correlation functions, such as the r.h.s.\ of~(\ref{decompB}),
very accurately over the full frequency range
for fairly large systems ($L\leq 128$) with
open boundary conditions 
and a {\sl finite\/} broadening factor $\eta$, i.e.,
the DDMRG actually gives
\begin{equation}
\sigma_{\eta;L}(\omega) =  \frac{1}{L} \sum_n 
\frac{ |\langle 0 | \hat{\jmath} | n\rangle|^2}{E_n-E_0}
\frac{\eta}{[\omega-(E_n-E_0)]^2+\eta^2} \; .
\label{convolution}
\end{equation}
For $\eta\to 0$, $\sigma_{\eta;L}(\omega)$ reduces to $\sigma_1(\omega)$
as defined in Eq.~(\ref{decomp}).
Ultimately, we are interested in the optical conductivity
of an infinite system, $L\to\infty$, for $\eta\to 0^+$.
It is shown in Ref.~\onlinecite{Eric2} that the most appropriate way of
approaching this double limit is to compute
$\sigma_{\eta;L}(\omega)$ for different system sizes while
keeping $\eta L=\hbox{const}$ and then to extrapolate to infinite
system size.
In this paper we use $\eta L= 12.8 t$, which yields an energy
resolution of $0.1 t$ for our largest system size ($L=128$).

A very useful consistency check of the method is to test various sum
rules, relating moments of the function $\sigma_1(\omega)$
to ground-state expectation values, which can be evaluated with great
accuracy using a standard DMRG method~\cite{Steve,DMRGbook}.
For instance, for the Hamiltonian~(\ref{Hamiltonian}) with
open boundary conditions
\begin{mathletters}
\label{sumr}
\begin{eqnarray}
\int_0^\infty \frac{d\omega}{\pi}
\omega\sigma_1(\omega)&=&
\frac{1}{L} \, \langle 0|\hat{\jmath}^2|0\rangle ,
\label{sumrA}\\[6pt]
\int_0^\infty \frac{d\omega}{\pi}
\sigma_1(\omega)&=&
\frac{e^2 t}{2L} \, \langle 0|\sum_{l;\sigma} 
\left( \hat{c}_{l,\sigma}^+\hat{c}_{l+1,\sigma} 
+ {\rm h.c.}\right) |0\rangle ,
\label{sumrB}\\[6pt]
\int_0^\infty \frac{d\omega}{\pi}
\frac{\sigma_1(\omega)}{\omega}&=&
\frac{e^2}{L} \, \langle 0|\Bigl[\sum_{l}
l(\hat{n}_l-1)\Bigr]^2|0\rangle\ .
\label{sumrC}
\end{eqnarray}
\end{mathletters}%
For $\sigma_{\eta;L}(\omega)$ these sum
rules are not fulfilled exactly, but only up to errors of
the order of ${\cal O}([\eta/t])$ or ${\cal O}([\eta/t]^2)$. 

The ground-state phase 
diagram of the Hamiltonian~(\ref{Hamiltonian}) exhibits several
different phases (for instance, Mott-Hubbard insulator, 
charge density wave, and bond-order wave for $U>V_1\geq 0$ 
and $V_2=0$, see Ref.~\onlinecite{EricUV}). 
To check the nature of the ground state for some fixed 
model parameters we calculate 
the spin gap and various ground state properties, such as charge 
density, bond order, and spin and density correlations, 
for large system sizes (up to
$L=512$ sites) with a standard DMRG method. 
The ground state of~(\ref{Hamiltonian})
is a Mott insulator for all values of the model parameters used 
in this work.  

With DMRG one can also calculate
the charge gap
\begin{equation}
E_{\rm c}(L) = E_0(L+1) + E_0(L-1) - 2 E_0(L) \, ,
\label{chargegap}
\end{equation}
where $E_0(N)$ is the ground state energy of~(\ref{Hamiltonian})
on a $L$-site lattice with $N$ electrons.
For $L\rightarrow\infty$, $E_{\rm c}$ gives the energy threshold 
of the electron-hole excitation continuum.
In a Mott insulator it corresponds to the Mott gap.
In the one-dimensional Hubbard model ($V_1=V_2=0$), it is
known that $E_{\rm c}$ is also equal to the optical gap 
which we define as the energy threshold of the lowest
band in the optical spectrum.
In all cases with $V_1, V_2 \neq 0$ discussed here we have found that 
the optical spectrum contains a single band,
which corresponds to unbound
particle-hole excitations, and that
$E_{\rm c}$ agrees with the onset of this band. 
Therefore, in this paper we identify the charge gap 
with the optical gap.   
(Some special cases for which 
the charge gap does not correspond to the optical gap
are discussed in Ref.~\onlinecite{EricUV}.)

We also use the symmetrized DMRG method~\cite{Rama2}
to calculate the
lowest eigenstates in the optical excitation symmetry 
subspace (see Sec.~\ref{Sec:model}). 
As the standard DMRG method, the symmetrized DMRG
yields not only the eigenenergies but also
allows us the computation of various expectation values and
correlation functions of the eigenstates
(for an example, see Sec.~\ref{Sec:excitonproperties}).
We can thus investigate the nature and properties
of these optical excitations. 
In particular, it is possible to distinguish unbound
particle-hole excitations from excitons and from
other kinds of excitations (excitonic strings, 
charge-density-wave droplets) which can dominate
the optical spectrum of a Mott insulator~\cite{EricUV}.
In this paper we consider only the regime of the 
Hamiltonian~(\ref{Hamiltonian}) where optically excited states
can be described as (bound or unbound) particle-hole pairs. 
We emphasize that the symmetrized DMRG results for
the optically excited states are always in perfect
agreement with the DDMRG results for the optical
conductivity, confirming the accuracy of both methods.

All DMRG methods have a truncation error which is 
reduced by increasing the number~$m$ of retained density matrix 
eigenstates 
(for more details, see Refs.~\onlinecite{Steve,DMRGbook}).  
Varying $m$ allows one to compute
physical quantities for different truncation errors and thus to obtain
error estimates on these quantities.
For some quantities, especially eigenenergies, it is possible 
to extrapolate the results to the limit of vanishing
truncation error and thus to achieve a greater accuracy.
We have systematically used these procedures to
estimate the precision of our numerical calculations
and adjusted the maximal number $m$ of density matrix 
eigenstates to reach a desired accuracy.
The largest number of density matrix eigenstates we
have used is $m=1200$.
For all numerical results presented in this paper 
DMRG truncation errors are negligible
unless specified explicitly.
The main cause of inaccuracies are finite size effects 
or extrapolation errors for $L\rightarrow\infty$ 
which we discuss below when we present our numerical results.

\subsection{Limit of large Mott gaps}
\label{Subsec:largegap}

Let us now consider the situation where
the Mott gap is much larger than the band width $4t$.
For large interaction strengths, $U\gg t,V_1,V_2$, it is possible to
analyze the model (\ref{Hamiltonian}) by means of a
$1/U$~expansion~\cite{Marburger}. 
If we ignore corrections of the
order $t/U$, all sites are singly occupied in the ground state.
Electron transfers are limited to processes that
conserve the number of double occupancies, and 
a rather simple band picture emerges for $V_1=V_2=0$.
In an optical absorption process we excite one hole 
at momentum~$k-q/2$ in the lower Hubbard band,
$\epsilon_{\rm LHB}(k)=-U/2+\epsilon(k)$, and one double occupancy 
at momentum~$k+q/2$ in the upper Hubbard band,
$\epsilon_{\rm UHB}(k)=U/2-\epsilon(k)$ (antiparallel bands). 
The total momentum of the two charge
excitations is~$q$, and their energy is~$\omega$.
Due to spin-charge separation, the oscillator strength
can be written as~\cite{Marburger}
\begin{equation}
|\langle 0 | \hat{\jmath} | n\rangle|^2
= \left|-ie \epsilon(k)\right|^2 g_q \; .
\end{equation}
The spin sector enters the current-current correlation function
via the momentum-dependent ground-state form factor $g_{q}$,
\begin{mathletters}
\begin{eqnarray}
&& g_q = 2 \langle 0 |
\hat{Z}_{r,r+1}^{+}(q)
\left(\frac{1}{4}-\hat{\rm\bf S}_r\hat{\rm\bf S}_{r+1} \right)
\hat{Z}_{r,r+1}^{\phantom{+}}(q) | 0 \rangle 
\\[6pt]
&& \hat{Z}_{r,r+1}^{\phantom{+}}(q) =
\frac{1}{L} \sum_l e^{-iql}
\hat{{\cal T}}_S^{(l-r)}\hat{{\cal T}}_{S'}^{-(l-r)}\; ,
\end{eqnarray}
\end{mathletters}%
where~$\hat{\cal T}_{S}$ shifts all spins by one site
whereas $\hat{\cal T}_{S'}$ performs the same operation
on the lattice with sites~$r$ and~$r+1$ removed.

For the large-$U$ Hubbard model itself,
the analysis of $g_q$ is rather involved. 
We can use a ``no-recoil approximation''~\cite{Marburger} to
argue that the dominant contributions to the conductivity come
from $q=0$ and $q=\pi$, which correspond to vertical
transitions between two antiparallel bands 
($q=0$; $\epsilon_{\rm LHB}(k)$, $\epsilon_{\rm UHB}(k))$
and between two parallel bands ($q=\pi$; 
$\epsilon_{\rm LHB}(k)$, $\epsilon_{\rm UHB}(k+\pi)$).
This hypothesis has subsequently been confirmed by
DDMRG~\cite{JGE,EricUV}, which yields excellent
agreement if the form factors are chosen
as $g_0 =2.65 $ and $g_{\pi}= 0.05 \pm 0.03$.
Exact sum rules impose the condition~$g_0+g_{\pi}=4\ln(2)\approx 2.77$
for an infinite system. The deviation of our best fits
can be traced back to finite-size effects and numerical
errors of the order of one percent.

We now discuss the effects of a finite nearest-neighbor
Coulomb repulsion~$V_1\ll U$ with $V_2=0$. 
It follows by direct inspection of the
Hamiltonian~(\ref{Hamiltonian}) that the double occupancy
and the hole now mutually attract. To some extent,
this is reminiscent of the situation encountered in Wannier
theory for a band insulator. However, unlike in Wannier theory,
the double occupancy and the hole are not fermionic quasi-particles
but spinless hard-core bosons. Even more importantly,
there is a critical value~$V_{\rm c}=2t$ below which no exciton
appears below the threshold of the particle-hole continuum 
at $\omega_{\rm ph}= E_{\rm c} = U-4t$. 
For $V_1>V_{\rm c}$ there is an exciton
at the energy
\begin{mathletters} 
\label{twoexcitonslargeU}
\begin{equation}
\omega_{1}= U-V_1-4t^2/V_1 \; .
\label{1stexc}
\end{equation}
In addition, there is a second Mott--Hubbard exciton at the energy
\begin{equation}
\omega_{2}= U-V_1\; ,
\label{2ndexc}
\end{equation}\end{mathletters}%
which, for $V_1<4t$, lies in the particle-hole
continuum but carries very little spectral weight.

The optical conductivity is given by
\begin{eqnarray}
\omega \sigma_1(\omega) & =& 
\pi g_{\pi} t^2 e^2 \delta(\omega-\omega_2)\nonumber\\
&&+  g_0 t^2 e^2 \biggl\{\! \Theta(V_1-2t) \pi \left(1-(2t/V_1)^2\right)
\delta(\omega-\omega_1)  \nonumber
\\
&&+ \Theta(4t-|\omega-U|)
\frac{2t^2\sqrt{1-((\omega-U)/4t)^2}}{V_1(\omega-\omega_1)}
\biggr\} \; . \nonumber\\
\label{largeUanalytic}
\end{eqnarray}
Here, $\Theta(x)$ is the Heaviside step function.
Apart from the two $\delta$-peaks corresponding to the excitons,
there is a particle-hole continuum for $|\omega -U|\leq 4t$.
Near the lower (upper) boundary, the optical conductivity
shows a characteristic square-root increase (decrease).
The only exception is the case of $V_1=V_{\rm c}$ where
the optical conductivity {\sl diverges\/} at the threshold,
\begin{equation}
\sigma_1(\omega) \propto \frac{1}{\sqrt{\omega-\omega_{\rm ph}}} \;
\; {\rm for } \; \; \omega- \omega_{\rm ph}\to 0^+ \quad (V_1=2t)\; .
\label{LutherEmerylargeU}
\end{equation}

The predictions of this strong-coupling analysis are 
confirmed by our numerical results.
Note that $\sigma_1(\omega)\sim 1/U$ so that it is more
convenient in DDMRG to calculate directly the imaginary part of the
current-current correlation function, i.e.,
${\rm Im}\left\{\chi(\omega)\right\}=\omega \sigma_1(\omega)$.
In analogy to Eq.~(\ref{convolution}), a broadening~$\eta$ is introduced
in the DDMRG procedure for the 
current-current correlation function,
\begin{equation}
{\rm Im}\left\{\chi_{\eta;L}(\omega)\right\} =  \frac{1}{L} \sum_n 
\frac{\eta \; |\langle 0 | \hat{\jmath} | n\rangle|^2}
{[\omega-(E_n-E_0)]^2+\eta^2}
\label{convolutionforchi} \; .
\end{equation}
For $\eta\to 0$, this expression reduces to
${\rm Im}\left\{\chi(\omega)\right\}$, and we analyze 
the finite-size effects as discussed above for
$\sigma_1(\omega)$.

\begin{figure}[ht]
\begin{center}
\epsfig{file=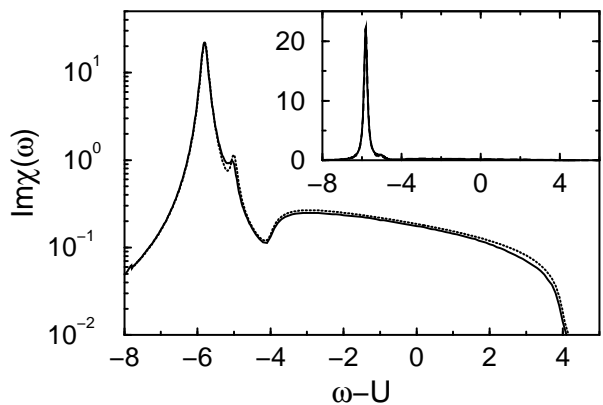,width=8cm}
\end{center}
\caption{Current-current correlation function 
${\rm Im}\left\{\chi(\omega)\right\}$ 
for $U/t \rightarrow \infty$, $V_1=5t$, $V_2=0$, and $\eta=0.1t$. 
The solid line is the DDMRG result 
${\rm Im}\left\{\chi_{\eta;L}(\omega)\right\}$
for $L=128$. The dotted line is Eq.~(\ref{largeUanalytic})
convolved with a Lorentzian of width $\eta$.
Note the logarithmic scale of the ordinate.
Inset: same results on a linear scale.}
\label{fig:HarrisLangewithV}
\end{figure}

Figure~\ref{fig:HarrisLangewithV} shows 
${\rm Im}\left\{\chi_{\eta,L}(\omega)\right\}$ 
for $L=128$ and $\eta=0.1 t$ obtained in the large-$U$ limit
of the extended Hubbard model with $V_1=5t$ and $V_2=0$.
We compare the DDMRG data to the analytical 
formula~(\ref{largeUanalytic}),
convolved with a Lorentzian of width~$\eta$.
The DDMRG data and the large-$U$ result are in very good agreement
when we choose $g_0=2.65$, $g_{\pi}=0.08$, as discussed above.
These values are found to be essentially independent of~$V_1$.
We previously obtained a similarly good agreement for the
case of the large-$U$ Hubbard model ($V_1=V_2=0$)~\cite{JGE}.

For $V_1=5t$, $V_2=0$, most of the spectral weight is carried by 
the exciton 
at $\omega-U=\omega_1-U = -5.8t$, 
as one also observes for typical band insulators.
Therefore we use a logarithmic scale to make visible 
the contributions of the other exciton and the particle-hole continuum.
The use of a logarithmic scale also reveals deviations 
around the second exciton ($\omega-U\approx\omega_2-U=-5t$)
and close to the upper band edge 
($\omega-U \approx +4t$),
which are associated with difficulties in the numerical
determination of small contributions to the optical spectrum,
and remaining finite-size and boundary effects.
In fact, these deviations are less than one percent
of the total spectral weight, and are completely irrelevant for
practical, i.e., experimental purposes. On a linear scale
they are not visible as seen in the  inset of 
Fig.~\ref{fig:HarrisLangewithV}.

The case of $0< V_2< V_1$
can be treated analogously and does not yield
any new physical aspects.
It is known that increasing the interaction range
simply reduces the critical coupling below
which no exciton appears in the gap and increases
the number of visible excitons in the spectrum~\cite{Gallagher}.

\subsection{Regime of intermediate Mott gaps}
\label{Subsec:intermediategap}

In the simple Hubbard model ($V_1=V_2=0$) we previously found
that the optical conductivity evolved smoothly from 
the regime of small Mott gaps ($U \ll t$)
to the limit of large Mott gaps ($U \gg t$)~\cite{JGE}.
Optical excitations can simply be interpreted
as a particle-hole pair, i.e., a pair of spinless
quasi-particles with opposite charges, corresponding
to the hole and double occupancy for large Mott gaps
and to the holon-antiholon
pair in the field-theory limit of small Mott gaps
(see Sec.~\ref{Sec:FT}).

The optical spectrum of the extended Hubbard model~(\ref{Hamiltonian})
shows more diversity in the presence of a finite
intersite Coulomb repulsion ($V_1 > V_2 \geq 0$).
Even within the Mott insulator phase one can observe ``exotic''
optical excitations such as charge-density-wave droplets
or excitonic strings~\cite{EricUV}.
In this study we restrict ourselves to the Mott insulator regime
of~(\ref{Hamiltonian}) where the dominant optical
excitations can be described as a (bound or unbound)
particle-hole pair and the Coulomb interaction is strong
enough to generate at least one bound pair (exciton).

Varying the model parameters $U/t$, $V_1/t$, and $V_2/t$
we have investigated the optical excitations of systems
with a Mott gap ranging from 10 to 0.1 times the band width $4t$.
It  is important to note that the Mott gap $E_{\rm c}$ increases
with $U$ and $V_2$ but decreases with increasing 
$V_1$~\cite{shuai,EricUV,Kancharla}.
As in the large Mott gap limit, we have found that the intersite
Coulomb interaction must exceed a critical value before a discrete
absorption peak appears at an energy $\omega_{\rm ex}$
below the optical gap $E_{\rm c}$.
For $V_2=0$ the critical value is $V_1\approx 2t$ for all $U/t$
in agreement with our analytical strong-coupling analysis and
a previous work~\cite{shuai2}.
The critical value of $V_1/t$ becomes smaller as the
next-nearest neighbor repulsion $V_2$ increases.

We have analyzed the nature of the excited states
associated with the discrete absorption peak
using various measurements and correlation functions.
For instance, in Sec.~\ref{Subsec:excitonsize} we present a method
to determine the size of a particle-hole pair.
This analysis confirms that this excited state is clearly
a bound particle-hole pair (exciton).
The exciton binding energies  $\delta E = E_{\rm c} - \omega_{\rm ex}$
observed in our calculations
range from  $\delta E=0.03t$ to $\delta E=12t$ and the exciton 
sizes measured with the procedure of Sec.~\ref{Subsec:excitonsize} vary
from $20 a_0$ down to slightly more than one lattice spacing.

It is interesting to note that we have never found more than one exciton
in the regime of intermediate Mott gaps.
Our numerical results (DDMRG and symmetrized DMRG) for
finite open chains sometimes yield more than one
optical excitations with energy
$\omega_q \approx \omega_{\rm ex} + c(L) q^2 < E_{\rm c}$ and 
quasi-momenta
$q \approx \pi (2\ell+1)/(L+1)$, 
$\ell=0,1,2,\ldots,\ell_{\rm max} \ll L/2$.~\cite{Robert}
The first of these states ($\ell=0$) has always much more
spectral weight than the other ones and
corresponds to an exciton with momentum $q=0$
in an infinite chain.
Scattering by the chain ends is responsible for
the small but finite optical weight of the
other states ($\ell \geq 1$), corresponding to
excitons with momentum $q \neq 0$ in
an infinite system. Thus, with periodic boundary conditions
or in an infinite system, only one exciton 
contributes to the optical conductivity
$\sigma_1(\omega)$ of the Hamiltonian~(\ref{Hamiltonian})
in the regime of intermediate Mott gaps.

In contrast to this, both the strong-coupling analysis
and field theory allow for more than one exciton
in the optical spectrum of a Mott insulator 
in the thermodynamic limit if
the Coulomb repulsion becomes strong enough. 
For the model~(\ref{Hamiltonian}) an increase of 
$V_1$ and $V_2$ does not lead to the formation
of a second Mott--Hubbard exciton. Instead, the nature
of the lowest optical excitations changes to
charge-density-wave droplets or excitonic strings,
or the ground state develops long-range order.
Both the strong-coupling analysis and the field-theory 
approach assume a Mott insulator ground state
and particle-hole pairs as optical excitations,
and thus do not reproduce this instability
toward charge density ordering.  
It is conceivable, though, that the inclusion of 
Coulomb terms beyond next-nearest neighbors
in the lattice model~(\ref{Hamiltonian})
favors the appearance of more excitons
in the optical conductivity of a Mott insulator.

Besides the single exciton peak we have always
found that $\sigma_1(\omega)$ shows an absorption band
starting at the charge gap $E_{\rm c}$. 
Within the resolution of our method
the optical spectrum does not display any other feature.
The investigation of the excited states in the continuum 
above $E_{\rm c}$
is much more demanding than the analysis of the isolated exciton peak.
Whenever this has been feasible, we have found 
that the excited states contributing to 
the absorption continuum can be described as an unbound
particle-hole pair (see Sec.~\ref{Subsec:excitonsize}).

\begin{figure}[ht]
\begin{center}
\epsfig{file=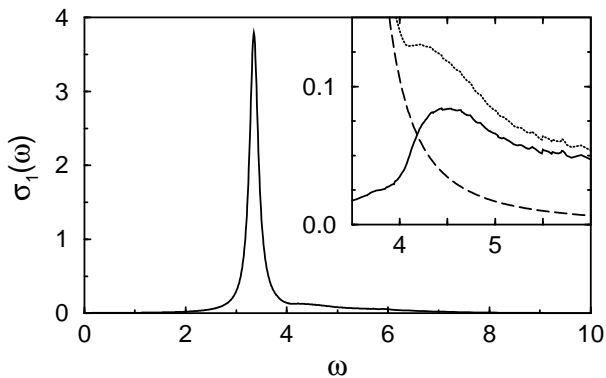,width=8cm}
\end{center}
\caption{Optical conductivity $\sigma_{\eta;L}(\omega)$
for $U=8t$, $V_1=4t$, and $V_2=2t$,
calculated with DDMRG on a 128-site lattice ($\eta=0.1t$).
The inset shows an expanded view of $\sigma_{\eta;L}(\omega)$
(dotted line) for $3.5 \leq \omega/t \leq 6$.
The exciton (dashed) and continuum (solid) contributions 
to $\sigma_{\eta;L}(\omega)$ are also shown.}
\label{fig:intermediate}
\end{figure}

As a typical example, the optical conductivity of~(\ref{Hamiltonian})
for $U=8t$, $V_1=4t$, and $V_2=2t$ is shown in 
Fig.~\ref{fig:intermediate}. 
A peak at the exciton energy $\omega_{\rm ex}=3.34t$ is the dominant
feature of the spectrum while a very weak band is visible
for $\omega \geq E_{\rm c} = 4.05t$.
No gap is visible between $\omega_{\rm ex}$ and $E_{\rm c}$ because
of the broadening of the strong exciton peak.
In the inset of Fig.~\ref{fig:intermediate}
one can see the weak particle-hole continuum part
of the spectrum separated from the strong exciton contribution.
The onset of the absorption band is clearly visible
at $\omega \approx E_{\rm c} = 4.05t$.
The small irregular fluctuations seen in the inset
are numerical errors (truncation errors)
made visible by the small scale used.

In summary, our numerical results show that
there is no qualitative change in the optical conductivity 
of a Mott insulator with excitons 
when one goes from the limit of large Mott gaps
down to the regime of intermediate Mott gaps with $E_{\rm c} 
\gtrsim 0.4t$.
As in the Hubbard model~\cite{JGE}, the simple picture of the 
strong-coupling analysis remains qualitatively valid. 

\section{Field Theory}
\label{Sec:FT}

In order to address the regime of small Mott gaps, we study the extended
Hubbard model~(\ref{Hamiltonian}) in the field-theoretical limit. This
limit can be constructed directly from the lattice model in the
weak-coupling regime $U,V_1, V_2\ll t$. The low-energy physics of the
non-interacting model is simply described in terms of a massless Dirac
fermion with velocity $v_{\rm F}=2ta_0$. The interactions introduce a
four-fermion coupling. The resulting effective theory is known as the
$U(1)$ Thirring model and can be represented as~\cite{GNT,CET2}
\begin{eqnarray}
{\cal H}&=& \int dx\left[{\cal H}_{\rm c}+{\cal H}_{\rm s}\right] ,
\nonumber\\
{\cal H}_{\rm s}&=& \sum_{\alpha}\left\{ \frac{2\pi v_{\rm s}}{3}\left[
:J^\alpha J^\alpha:+:\bar{J}^\alpha \bar{J}^\alpha:\right]
-2g_\perp\ {J}^\alpha\bar{J}^\alpha\right\}\, ,\nonumber\\
{\cal H}_{\rm c}&=& \sum_{\alpha} 
\frac{2\pi v_{\rm c}}{3}\left[
:I^\alpha I^\alpha:+:\bar{I}^\alpha \bar{I}^\alpha:\right]
\nonumber\\
&&\hphantom{\sum_{\alpha}\biggl\{ }
+\left[{g_\perp}
\left(I^x\bar{I}^x+I^y\bar{I}^y\right)+g_\parallel\ I^z \bar{I}^z 
\right] \, ,
\label{HU1}
\end{eqnarray}
where $J^\alpha$, $\bar{J}^\alpha$ ($I^\alpha$, $\bar{I}^\alpha$) are
left and right moving SU(2) spin currents (SU(2) pseudospin currents) 
and
\begin{eqnarray}
g_{\perp}&=&2(U-2[V_1-V_2])a_0\; ,\;
g_{\parallel}=2(U+6V_1+2V_2)a_0\; ,\nonumber\\[6pt]
v_{\rm c}&=& v_{\rm F}+\frac{2(U/4+V_1+V_2)a_0}{\pi} \;,\;
v_{\rm s}=v_{\rm F}-\frac{Ua_0}{2\pi}\; .
\label{gfactors}
\end{eqnarray}
The Hamiltonian (\ref{HU1}) explicitly exhibits spin-charge
separation: ${\cal H}_{\rm c,s}$ describe the 
charge and spin degrees of
freedom, respectively, which are independent of one another. As long as
$g_\perp>0$, the spin sector remains gapless and can be bosonized in
terms of a Gaussian model. The charge sector can be bosonized as well,
as is for example shown in Ref.~\onlinecite{GNT}. The result is a
Sine-Gordon model (SGM)
\begin{eqnarray}
{\cal H}_{\rm c}&=&\frac{1}{16\pi}
\left[(\partial_t\phi_{\rm c})^2+(v_{\rm c}\partial_x\phi_{\rm c})^2
\right]
+2\mu_{\rm c}\cos\beta\phi_{\rm c}\ ,
\label{HamiltFT}
\end{eqnarray}
where $\mu_{\rm c}$ and $\beta$ are functions of $U$, $V_1$, $V_2$.
In order to utilize results obtained from the integrability of the
Sine-Gordon model we use a flat renormalization scheme, in which
$\beta$ is a constant
\begin{equation}
\beta^2=\frac{4\pi v_{\rm F}}{4\pi v_{\rm F}+
\sqrt{g_{\parallel}^2-g_{\perp}^2}}\; .
\label{betasmallcouplings}
\end{equation}
The pure Hubbard model corresponds to the limit $\beta\to 1$ and the 
effect of
$V_1$ and $V_2$ is to decrease the value of $\beta$.

In the field-theory limit the electrical current operator 
is found to be
\begin{equation}
{\cal J}(t,x) = \sqrt{A^{\prime}} \partial_t \phi_{\rm c} \; ,
\label{currentFT}
\end{equation}
where $A^{\prime}>0$ is a non-universal constant.
As seen from Eq.~(\ref{currentFT}), the current operator
does not couple to the spin sector. This shows that spinons do not
contribute to the optical conductivity. Therefore we ignore
the spin part ${\cal H}_{\rm s}$ of the Hamiltonian from now on.

The calculation of the optical conductivity in the field-theory
limit has thus been reduced to the evaluation of the
retarded current-current
correlation function in the charge sector,
\begin{equation}
\chi^{\rm FT}(\omega)=\frac{i}{La_0}\int_0^{\infty}\!\! dt \exp(i\omega t)
\int_{-\infty}^{\infty}\!\! dx
\langle [{\cal J}(x,t), {\cal J}(0,0)]\rangle .
\label{chiFT}
\end{equation}
We turn to the calculation of this correlation function
in Sec.~\ref{Subsec:sine-Gordon}.

For the Sine-Gordon model, many exact results are
available~\cite{SG}. The spectrum of the SGM
depends on the value of the coupling constant~$\beta$.
In the so-called repulsive regime, $1/\sqrt{2}<\beta<1$,
there are only two elementary excitations, called soliton
and antisoliton. From the point of view of the underlying lattice
model~(\ref{Hamiltonian}) these correspond to holon and antiholon
(spinless excitations of opposite charges).
These have a massive relativistic dispersion,
\begin{equation}
E(P)=\sqrt{M^2+v_{\rm c}^2 P^2} \; ,
\label{holondispersion}
\end{equation}
where $M$ is the single-particle gap which is related to
the optical gap by~$\Delta=2M$. At weak coupling the gap scales as
\begin{equation}
M\approx\frac{8t}{\sqrt{2\pi}}
\sqrt{g(1+x)}\left(\frac{1-x}{1+x}\right)^{(gx+2)/{4gx}}\ ,
\label{FTgap1}
\end{equation}
where we have fixed the constant factor by comparing to the exact 
result 
for the Hubbard model, and where
\begin{eqnarray}
x&=&\left[1-\left(\frac{U-2V_1+2V_2}{U+6V_2+2V_2}\right)^2\right]^{1/2}
\ ,\nonumber\\[6pt]
g&=&(U+6V_1+2V_2)/2\pi t\ .
\label{FTgap2}
\end{eqnarray}
We note that the gap vanishes on the critical surface
$U-2V_1+2V_2=0$ separating the Mott-insulating phase
with gapless spin excitations from another phase
with a spin gap.

In the regime $0<\beta <1/\sqrt{2}$, soliton and antisoliton
attract and can form bound states. In the SGM these are usually
known as ``breathers'' and correspond to excitons in our lattice
model. There are 
\begin{equation}
N_{\rm ex}= \left[ \frac{1-\beta^2}{\beta^2}\right]
\label{Nex}
\end{equation}
different types of excitons, where $[ x ]$ in~(\ref{Nex}) denotes
the integer part of~$x$.
The exciton gaps are given by
\begin{equation}
M_n=2M \sin(n\pi\xi/2)\ ,\quad n=1,\ldots ,N_{\rm ex}\; ,
\label{exgap}
\end{equation}
where 
\begin{equation}
\xi=\frac{\beta^2}{1-\beta^2} \; .
\label{xi}
\end{equation}
Therefore, the single-particle gap~$M$
and the coupling~$\beta$ fully
characterize the spectrum of the SGM.

One knows that the field-theory approximation
to the lattice problem is valid
in the limit $U,V_1,V_2\ll t$ where the
single-particle gap is much smaller than the bandwidth.
For the Hubbard model ($V_1=V_2=0$) we have found~\cite{JGE}
that field theory gives surprisingly good
results for the optical conductivity even for intermediate
single-particle gaps of magnitude $M\approx 0.3 t$. The same holds true
for the Mott insulating phase of
the extended Hubbard model with $V_1>0$, $V_2=0$~\cite{EricUV,CET}.
As we shall show in Sect.~\ref{Subsec:FTtoDMRG}, field theory 
remains applicable even in the presence of excitons.

In the framework of the field-theory approximation
to the lattice problem we can determine the value of~$\beta$
only in the limit $U,V_1,V_2\ll t$.
In fact, for the Hubbard model $\beta=1$ is fixed by the
${\rm SO}(4)$~symmetry~\cite{SO4}. As seen from
Eq.~(\ref{betasmallcouplings}), the effect of
a small $V_1$, $V_2$ is to decrease the value of $\beta$.
One may therefore hope that by carefully tuning $U$, $V_1$ and $V_2$
one may stay in a regime with a ``small'' single-particle gap,
i.e., close to the critical surface $U-2V_1+2V_2=0$, but with
a sufficiently small $\beta$ for excitons to exist.
We have found numerically that it is indeed possible to reach regions
of the parameter space
where field theory is valid and $\beta^2$ is
as small as $0.36$ close to a critical
surface $U-2V_1+2V_2 \approx 0$
which separates the Mott insulating phase from 
other phases with long-range order.
The determination of the field-theoretical parameter
$\beta^2$ as a function of the lattice model
parameters using DMRG results
will be discussed in Sec.~\ref{Subsec:FTtoDMRG}.

\subsection{Optical conductivity in the Sine-Gordon model}
\label{Subsec:sine-Gordon}

Our task is now to calculate the Fourier transform of the retarded,
dynamical current-current correlation function~(\ref{chiFT}) in the
Sine-Gordon model. This is done by going to the spectral
representation and then utilizing the integrability of the
SGM to determine exactly the matrix elements of the current operator
between the ground state and various excited states. This method is
known as the form-factor bootstrap approach \cite{smirnov,FF} and has
recently been applied to calculate the optical conductivity in the
repulsive regime of the SGM \cite{JGE,CET}. Here we review briefly the
relevant steps and refer to~Ref.~\onlinecite{CET2} 
for a more detailed exposition.
In order to utilize the spectral representation we need to specify a
basis of eigenstates of the Hamiltonian (\ref{HamiltFT}). Such a basis
is given by scattering states of excitons, holons and antiholons. In
order to distinguish these we introduce labels $e_1,e_2,\ldots
e_{N_{\rm ex}},h,\bar{h}$. As usual for particles with relativistic
dispersion, it is useful to introduce a rapidity variable $\theta$ to
parameterize energy and momentum
\begin{mathletters}
\begin{eqnarray}
E_{h}(\theta)&=&M\cosh\theta\; , \; P_{h}(\theta)=(M/v_{\rm c})
\sinh\theta\; ,\\[3pt]
E_{\bar{h}}(\theta)&=&M\cosh\theta\; , \; 
P_{\bar{h}}(\theta)=(M/v_{\rm c})\sinh\theta \; ,\\[3pt]
E_{e_n}(\theta)&=&M_n\cosh\theta\; , \; 
P_{e_n}(\theta)=(M_n/v_{\rm c})\sinh\theta\; , 
\end{eqnarray}\end{mathletters}%
where the exciton gaps $M_n$ are given by~(\ref{exgap}). Next we
turn to the construction of a basis of scattering states of holons,
antiholons and excitons. A convenient formalism to this end is 
is obtained by using the Zamolodchikov-Faddeev (ZF) algebra. The ZF
algebra can be considered to be the extension of the algebra
of creation and annihilation operators for free fermions or bosons to
the case or interacting particles with factorizable scattering.
The ZF algebra is based on the knowledge of the exact spectrum and
scattering matrix of the model \cite{SM}. For the SGM
the ZF operators (and their hermitian
conjugates) satisfy the following algebra
\begin{mathletters}
\label{fz1}
\begin{eqnarray}
{Z}^{\epsilon_1}(\theta_1){Z}^{\epsilon_2}(\theta_2) &=&
S^{\epsilon_1,\epsilon_2}_{\epsilon_1',\epsilon_2'}(\theta_1 -
\theta_2){Z}^{\epsilon_2'}(\theta_2){Z}^{\epsilon_1'}(\theta_1)\; ,
\\
{Z}_{\epsilon_1}^\dagger(\theta_1)Z_{\epsilon_2}^\dagger(\theta_2) &=&
Z_{\epsilon_2'}^\dagger(\theta_2){ Z}_{\epsilon_1'}^\dagger
(\theta_1)S_{\epsilon_1,\epsilon_2}^{\epsilon_1',\epsilon_2'}(\theta_1 -
\theta_2)\; ,\\
Z^{\epsilon_1}(\theta_1)Z_{\epsilon_2}^\dagger(\theta_2) 
&=&Z_{\epsilon_2'}
^\dagger(\theta_2)
S_{\epsilon_2,\epsilon_1'}^{\epsilon_2',\epsilon_1}
(\theta_2-\theta_1)Z^{\epsilon_1'}(\theta_1)\nonumber\\
&&+(2 \pi) \delta_{\epsilon_2}^{\epsilon_1} 
\delta (\theta_1-\theta_2)\; .
\end{eqnarray}\end{mathletters}%
Here $S^{\epsilon_1,\epsilon_2}_{\epsilon_1',\epsilon_2'}(\theta)$ are
the known (factorizable) two-particle scattering matrices \cite{SM} and
$\varepsilon_j=h,\bar{h},e_1,\ldots,e_{[1/\xi]}$.

Using the ZF operators a Fock space of states can be constructed as
follows. The vacuum is defined by
\begin{equation}
Z_{\varepsilon_i}(\theta) |0\rangle=0 \ .
\end{equation}
Multiparticle states are then obtained by acting with strings of
creation operators $Z_\epsilon^\dagger(\theta)$ on the vacuum
\begin{equation}
|\theta_n\ldots\theta_1\rangle_{\epsilon_n\ldots\epsilon_1} = 
Z^\dagger_{\epsilon_n}(\theta_n)\ldots
Z^\dagger_{\epsilon_1}(\theta_1)|0\rangle . 
\label{states}
\end{equation} 
In term of this basis the resolution of the identity is given by
\begin{eqnarray}
&& \openone = |0\rangle\langle 0| \label{identity} \\
 && + \sum_{n=1}^\infty\sum_{\epsilon_i}\int_{-\infty}^{\infty}
\frac{d\theta_1\ldots d\theta_n}{(2\pi)^nn!}
|\theta_n\ldots\theta_1\rangle_{\epsilon_n\ldots\epsilon_1}
{}^{\epsilon_1\ldots\epsilon_n}\langle\theta_1\ldots\theta_n|\ .
\nonumber
\end{eqnarray}
Inserting (\ref{identity}) between the current operators in~(\ref{chiFT}), 
we obtain the following spectral representation of the
correlation function
\begin{eqnarray}
\label{corr}
&&\langle {\cal J}(x,t){\cal J}(0,0)\rangle
=\sum_{n=1}^\infty\sum_{\epsilon_i}\int
\frac{d\theta_1\ldots d\theta_n}{(2\pi)^nn!}
\nonumber\\
&&\times
\exp\Bigl({i\sum_{j=1}^n P_jx-E_j t}\Bigr)
|\langle 0| {\cal J}(0,0)|\theta_n\ldots\theta_1
\rangle_{\epsilon_n\ldots\epsilon_1}|^2, 
\end{eqnarray}
where 
$P_j$ and $E_j$ are given by
\begin{equation}
P_j=\frac{M_{\epsilon_j}}{v_{\rm c}}\sinh \theta_j, \;
E_j=M_{\epsilon_j}\cosh \theta_j\ ,
\label{epj}
\end{equation}
and 
\begin{equation}
\label{formf}
f^{\cal J}(\theta_1\ldots\theta_n)_{\epsilon_1\ldots\epsilon_n}\equiv
\langle 0| {\cal
J}(0,0)|\theta_n\ldots\theta_1\rangle_{\epsilon_n\ldots\epsilon_1} 
\end{equation}
are the form factors. Our conventions in (\ref{epj}) are such that
$M_h=M_{\bar{h}}=M$ and $M_{e_n}=M_n$. 
After carrying out the double Fourier transform we arrive at
\begin{eqnarray}
\label{expansion1}
&&\chi^{\rm FT}(\omega)= 
\frac{-2\pi}{La_0}\sum_{n=1}^\infty\sum_{\epsilon_i}\!\int\!
\frac{d\theta_1\ldots d\theta_n}{(2\pi)^nn!}
|f^{\cal J}(\theta_1\ldots\theta_n)_{\epsilon_1\ldots\epsilon_n}|^2 
\nonumber\\
&&\times 
\Bigl[\frac{\delta(\sum_jM_{\epsilon_j}\sinh\theta_j/v_{\rm c})}
{\omega - \sum_j
M_{\epsilon_j}\cosh\theta_j +i\eta}-
\frac{\delta(\sum_jM_{\epsilon_j}\sinh\theta_j/v_{\rm c})}
{\omega + \sum_j
M_{\epsilon_j}\cosh\theta_j +i\eta}\Bigr] .\nonumber\\
\end{eqnarray}
This then yields the following representation for the
real part of the optical conductivity ($\omega>0$)
\begin{eqnarray}
\sigma_1^{\rm FT}(\omega)&=& \frac{2 \pi^2}{La_0\omega}
\sum_{n=1}^\infty\sum_{\epsilon_i}\int 
\frac{d\theta_1\ldots d\theta_n}{(2\pi)^n n!}\nonumber\\
&\times& \left |
f^{\cal J}(\theta_1\ldots\theta_n)_{\epsilon_1\ldots\epsilon_n}
\right | ^2 \label{expansion2} \\
 &\times&\delta(\sum_k\frac{M_{\epsilon_k}}{v_{\rm c}}\sinh\theta_k)
\delta(\omega - \sum_k 
M_{\epsilon_k}\cosh\theta_k)\; . \nonumber 
\end{eqnarray}
The missing ingredient in~(\ref{expansion2}) are the form factors.
In Refs.~\onlinecite{smirnov,FF} 
integral representations for the form factors of 
the
current operator in the Sine-Gordon model were derived. Using these
results we can determine the first few terms of the 
expansion~(\ref{expansion2}). In particular, the form factors 
involving excitons are determined via the bootstrap axioms for 
soliton-antisoliton form factors \cite{smirnov}. 

{}From (\ref{expansion2}) it is easy to see for any given frequency
$\omega$ only a finite number of intermediate states will contribute:
the delta function forces the sum of single-particle gaps
$\sum_jM_{\epsilon_j}$ to be less than $\omega$.
Expansions of the form (\ref{expansion2}) are usually found
to exhibit a rapid convergence, which can be understood in terms of
phase space arguments~\cite{oldff1,mussardo.school}. Therefore we
expect that summing the first few terms in~(\ref{expansion2}) will
give us good results over a large frequency range.

Using the transformation property of the current operator under charge
conjugation one finds that many of the form factors in
(\ref{expansion1}) actually vanish. In particular, only the ``odd'' 
excitons 
$e_1, e_3,\ldots$ (assuming they exists, i.e., $\beta$ is sufficiently
small) couple to the current operator. The first few non-vanishing terms
of the spectral representation~(\ref{expansion2}) are
\begin{equation}
\frac{\sigma_1^{\rm FT}(\omega)}{A} = 
\sum_{n=1}^{[(N_{\rm ex}+1)/2]}\!\!\!\! \sigma_{e_{2n-1}}(\omega)
+\sigma_{h\bar{h}}(\omega)
+\sigma_{e_1e_2}(\omega)+\ldots
\label{sigmaSR}
\end{equation}
Here $A=A^{\prime} v_{\rm c}/ La_0$ and $\sigma_{e_n}(\omega)$,
$\sigma_{h\bar{h}}(\omega)$ and $\sigma_{e_1e_2}(\omega)$ are the
contributions of the odd excitons, the holon-antiholon continuum and
the $e_1e_2$ exciton-exciton continuum respectively. The latter of
course exists only if $N_{\rm ex}\geq 2$. 
We find
\begin{mathletters}
\begin{eqnarray}
&&\sigma_{e_n}(\omega)=
\frac{\pi}{M_{2n-1}^2}f_{2n-1}\delta(\omega -
M_{2n-1})
\\
&&f_{m}=4M^2\xi^2\ \sin(m\pi\xi)\prod_{n=1}^{m-1}\tan^2(\pi
n\xi/2)\\
&&\times\exp\left(-2\int_0^\infty
\frac{dt}{t}\frac{\sinh(t(1-\xi)/2)}{\sinh(t\xi/2)\cosh(t/2)}
\frac{\sinh^2(mt\xi/2)}{\sinh t}\right)\; .\nonumber
\end{eqnarray}
\label{excitonsigma}
\end{mathletters}%
The holon-antiholon contribution has previously been 
determined~\cite{CET} 
and is given by
\begin{eqnarray}
&&\sigma_{h\bar{h}}(\omega)=
\frac{4\sqrt{\omega^2-4M^2}\Theta(\omega-2M)}
{\omega^2[\cos({\pi}/{\xi})+\cosh(\theta/\xi)]}
\label{2part}\label{continuumsigma}\\
&&\times\exp\left(\int_0^\infty
\frac{dt}{t}\frac{\sinh[t(1-\xi)/2]\left[1-\cos(t\theta/\pi)\cosh t
\right]
}{\sinh(t\xi/2)\cosh(t/2)\sinh t}
\right)\, ,
\nonumber
\end{eqnarray}
where $\theta=2 {\rm arccosh} (\omega/2M)$.
Finally, we quote the result for the $e_1e_2$ exciton-exciton continuum
\begin{eqnarray}
\sigma_{e_1e_2}(\omega)=
\frac{2\omega|f_{12}(\theta_{12})|^2}{\sqrt{(\omega^2-M_1^2-M_2^2)^2
-(2M_1M_2)^2}}\, ,
\end{eqnarray}
where
\begin{eqnarray}
&&|f_{12}(\theta_{12})|^2=\lambda^6\frac{|\tan\pi\xi|}{2}
\frac{\sinh^2\theta_{12}+\sin^2(\pi\xi/2)}
{\sinh^2\theta_{12}+\sin^2(3\pi\xi/2)}
\nonumber\\
&&\times\exp\left(\!\!-8\int_0^\infty\!\frac{dt}{t}\frac{\sinh t 
\sinh (t\xi)
\sinh[t(1+\xi)] \cosh (2t\xi)}{\sinh^2(2t)}\right)\nonumber\\
&&\times 
\exp\left(\!\!-4\int_0^\infty\!\frac{dt}{t}\frac{\sinh (2t\xi) 
\sinh [t(1+\xi)] \cos (2t\theta_{12}/\xi)}{\cosh t \sinh(2t)}\right) 
\nonumber \\
&& \times
\left(4\cos\left[2\pi(\cosh\theta_{12}+\cos(\pi\xi/2))/\xi\right]
\right)^{-2}
\; ,
\end{eqnarray}
and
\begin{eqnarray}
\theta_{12}&=&{\rm arccosh}
\left(\frac{\omega^2-M_1^2-M_2^2}{2M_1M_2}\right)\ ,\\
\lambda&=&2 \cos\left(\frac{\pi\xi}{2}\right)\sqrt{2\sin
\left(\frac{\pi\xi}{2}\right)}
\exp\left(-\int_0^{\pi\xi}\frac{dt}{2\pi}\frac{t}{\sin t}\right)\; .
\nonumber 
\end{eqnarray}

For $1 \geq \beta^2 > 1/3$ 
the contributions
of the first odd exciton~(\ref{excitonsigma})
and of the holon-antiholon continuum~(\ref{continuumsigma}) 
dominate the field-theoretical optical conductivity~(\ref{sigmaSR})
(at least in the low-frequency regime)
and other contributions vanish or can be neglected. 
The optical conductivity can 
be written as  
\begin{equation}
\sigma_1^{\rm FT}(\omega) = \frac{A}{\Delta} 
h_{\beta}\left (\frac{\omega}{\Delta} \right ) \; ,
\label{universalsigma}
\end{equation}
where $h_{\beta}(x)$ is a universal function
depending only on $\beta^2$.
Therefore, only the parameter $\beta^2$ determines the shape of
the optical spectrum. The parameters $\Delta$
and $A/\Delta$  just set the energy and conductivity scales,
respectively.

For $1 \geq \beta^2 \geq 1/2$ the optical spectrum
contains a single band, while for
$1/2 > \beta^2 > 1/3$
it contains
one band and one exciton peak in the optical gap 
at the energy $\omega_{\rm ex} = M_1$. 
The optical weight is progressively transfered
from the band to the single exciton as
$\beta^2$ decreases down to $1/3$.
The absorption band increases smoothly at the
threshold $\Delta$, as
\begin{equation}
\sigma_1^{\rm FT}(\omega) \propto \sqrt{\omega - \Delta} \quad \hbox{for}
\quad \omega \to \Delta^{+} \; ,
\end{equation}
for all values of $\beta^2$ but $\beta^2 = 1/2$, 
 when the exciton peak merges with the band.
In this case $\sigma_1^{\rm FT}(\omega)$ shows a square-root divergence
at the absorption threshold 
\begin{equation}
\sigma_1^{\rm FT}(\omega) \propto \frac{1}{\sqrt{\omega - \Delta}}
\quad \hbox{for} \quad
\omega \to \Delta^{+} \; (\beta^2=1/2)\; .
\end{equation}
The behavior of the field-theoretical conductivity
is thus qualitatively similar to that of the
lattice model in the large Mott-gap 
limit~(\ref{largeUanalytic}).
For $V_2 =0$~\cite{EricUV} it has also been found
for intermediate Mott gaps that 
$\sigma_1(\omega)$ diverges as a square-root
at the absorption band threshold $E_{\rm c}$ for the critical
coupling $V=V_{\rm c}$ 
below which no exciton appears in the optical gap.
For $V_1 \neq V_{\rm c}$, 
$\sigma_1(\omega)$ increases      
smoothly at the absorption band threshold.
This generic behavior of $\sigma_1(\omega)$
illustrates once more the absence of significant
qualitative changes in the particle-hole continuum
and single-exciton spectrum of (\ref{Hamiltonian})
as one goes from the large to the small Mott gap limit.
For $\beta^2 \leq 1/3$, the field-theoretical
optical conductivity shows more features, but it seems
that this regime cannot be reached in the lattice 
model~(\ref{Hamiltonian}) and thus we shall not discuss it further.

\subsection{Application to the lattice model}
\label{Subsec:FTtoDMRG}

The field theory optical conductivity
$\sigma_1^{\rm FT}(\omega)$, (Eq.~\ref{sigmaSR}),
depends on three parameters: $\beta^2$, the gap $\Delta=2M$,
and the normalization $A$. 
To compare the field theory predictions with our 
numerical results for the lattice model~(\ref{Hamiltonian})
one needs to determine the field theory parameters corresponding 
to specific values of the model parameters 
$U$, $V_1$, and $V_2$.
Fortunately, if there is exactly one exciton in the spectrum
($1/2 > \beta^2 > 1/3$), this can be done
using standard DMRG techniques.

The first step is the calculation of the gap parameter $\Delta$.
The gap $\Delta$ obviously corresponds to the charge 
gap~(\ref{chargegap}) extrapolated to the infinite system limit,
$\Delta = \lim_{L\rightarrow \infty} E_{\rm c}(L)$.
The second step is determining $\beta^2$.
The exciton energy $\omega_{\rm ex}(L)$ of the lattice system
can be calculated using the symmetrized DMRG method.
The extrapolation to infinite system size gives us the
exciton gap of field theory, Eq.~(\ref{exgap}), 
$M_1 = \lim_{L\rightarrow \infty} \omega_{\rm ex}(L)$.
The parameter $\beta^2$ is then fixed by the ratio of $M_1$
and $\Delta=2M$ through Eqs.~(\ref{exgap}) and~(\ref{xi}).

The final step is the calculation of the normalization~$A$.
Evaluating $A$ accurately turns out to be the most
difficult task and we have tried two different approaches.
In the first one we calculate the excitonic spectral weight
in the lattice model using the symmetrized DMRG technique
and extrapolate to
an infinite system size. This yields the normalization
$A$ by comparison with the field-theory prediction, 
Eqs.~(\ref{sigmaSR}) and~(\ref{excitonsigma}).
This method is simple and exact on the field theory
side but difficult to carry out numerically
because of significant truncation errors
and complicated finite-size effects. 

In the second approach we use the sum rules~(\ref{sumr}).
We numerically integrate the contributions of
the exciton and the holon-antiholon continuum to the
optical conductivity~(\ref{sigmaSR}) for a fixed value of $\beta^2$.
The result has a trivial dependence on $A$ and $\Delta$
because of~(\ref{universalsigma}).
This gives us the l.h.s.~of~(\ref{sumrB}) or~(\ref{sumrC})
assuming that the neglected contributions to $\sigma_1^{\rm FT}(\omega)$
are insignificant and that most of the optical
spectrum weight is concentrated at low energy where
field theory is expected to describe the
lattice-model properties accurately.
The r.h.s.~of~(\ref{sumrB}) and~(\ref{sumrC}) can be calculated
for the lattice model with DMRG and extrapolated to the
infinite system limit. Comparison of both sides of the
equations give the value of the normalization $A$.

We prefer the second approach because it is simpler
and more accurate than the first one as far as the numerical
calculations are concerned. It can also be used when
there is no exciton ($1 \geq \beta^2 \geq 1/2$).
Unfortunately, the second approach works only if the 
exciton~(\ref{excitonsigma}) and holon-antiholon 
continuum~(\ref{continuumsigma}) 
reproduce the lattice model $\sigma_1(\omega)$
very accurately, i.e., if
\begin{equation}
\delta S_n = \int_{0}^{\infty} d\omega \; \omega^{-n} \; 
|A \, \sigma_{e_{1}}(\omega) 
+ A \, \sigma_{h\bar{h}}(\omega) - \sigma_1(\omega)| 
\end{equation}
is very small compared to 
the l.h.s~of~(\ref{sumrB}) for $n=0$ 
or the l.h.s.~of~(\ref{sumrC}) for $n=1$.

The validity of the field-theory approximation to the
model~(\ref{Hamiltonian}) is not necessarily restricted to the
limit $U,V_1,V_2 \ll t$ but rather to the regions where
the single-particle gap $M=\Delta/2=E_{\rm c}/2$ is small compared to 
the
band width $4t$.
In the Hubbard model~\cite{JGE}
($V_1=V_2=0$) we have found that field theory describes
the optical conductivity accurately even for $U=3t$,
corresponding to $\Delta \approx 0.6t$.
In the case $V_1 > 0$, $V_2=0$, field theory is valid not only
in the weak-coupling limit ($U, V_1 \ll t$) but also
in the vicinity of a critical line $U-2V_1 \approx 0$
separating the
Mott-insulating phase from phases with long-range order~\cite{EricUV}.
The Mott gap vanishes or becomes extremely small on this critical
line at least up to $U=4t$.
The holon-antiholon continuum
contribution~(\ref{continuumsigma})
to the optical conductivity agrees  with
DDMRG results for
gaps $\Delta \approx 0.6t$ and $\beta^2$
ranging from 1 to $1/2$.

In the general case ($U > V_1 > V_2 > 0$), there is also 
a critical surface $U-2V_1+2V_2 \approx 0$ separating the
Mott insulating phase from other phases (in particular,
a charge-density-wave phase), where the Mott gap
vanishes or becomes very small even for relatively strong
Coulomb interaction (at least up to $U=6t$). 
A careful tuning of the parameters $U$, $V_1$, and $V_2$
allows one to reach regions of the parameter space,
where field theory appears to remain valid and $\beta^2$ decreases
down to 0.36 (for instance, $U=6t$, $V_1=4.5t$, and $V_2=2t$).
We can thus compare our numerical results with the
field-theoretical predictions for the optical
conductivity in the parameter regime with a single exciton.
Again, we have found a good agreement between
DDMRG and field theory for gaps as large as $\Delta=0.6t$. 

\begin{figure}[t]
\begin{center}
\epsfig{file=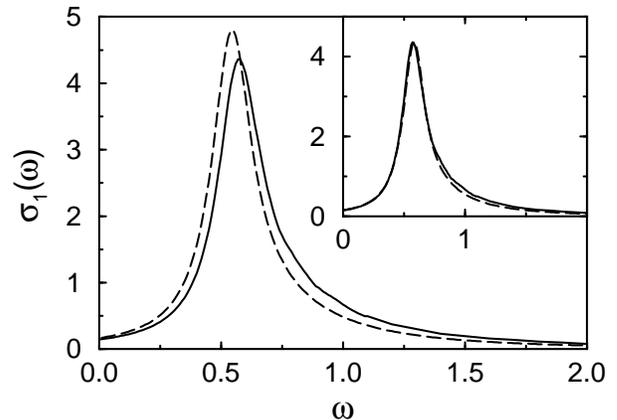,width=8cm}
\end{center}
\caption{Optical conductivity $\sigma_{\eta;L}(\omega)$
(solid) calculated with DDMRG for $U=5.2t$, $V_1=3.7t$, 
and $V_2=1.3t$ on a 128-site lattice ($\eta=0.1t$).
Field-theoretical result $\sigma_1^{\rm FT}(\omega)$ (dashed)
for $\Delta = 0.625t$, $\beta^2 = 0.40$ and $A=1.12e^2t$, 
convolved with a Lorentzian of width $\eta=0.1t$.
Inset: same results with $\Delta = 0.669t$ and
$A=1.02e^2t$. }
\label{fig:DMRGvsFT}
\end{figure}

In Fig.~\ref{fig:DMRGvsFT} we compare
the optical conductivity from field theory and DDMRG 
for $U=5.2t$, $V_1=3.7t$, and $V_2=1.3t$.
Only the low-frequency ($\omega \leq 2t$)
part of the spectrum is shown as there is almost no spectral weight at
higher frequency.
The numerical results have been calculated on a 128-site chain
with a resolution of $\eta=0.1t$.
The field-theory result,
Eqs.~(\ref{sigmaSR})--(\ref{continuumsigma}), 
has been convolved with a Lorentzian
of width $\eta=0.1t$ to allow for
a direct comparison with DDMRG results.
The field theory parameters $\Delta=0.625t$, $\beta^2=0.40$, 
and $A=1.12 e^2t$ have been evaluated for an infinite chain
as discussed above.
The gap between the exciton peak at $M_1=0.544t$ 
and the threshold of a weak 
holon-antiholon continuum at $\Delta$
is not visible in Fig.~\ref{fig:DMRGvsFT} because
it is smaller than the finite resolution introduced
by the broadening. 

In Fig.~\ref{fig:DMRGvsFT} one sees that the agreement
between numerical and field-theory results is good.
The visible discrepancies are well-understood finite-size effects:
the excitation energies of (\ref{Hamiltonian}),
in particular the exciton energy,
decreases as the system size increases,
the total spectral weight is slightly smaller in the
finite chain as shown by corrections to the 
kinetic energy (r.h.s.~of~(\ref{sumrB}))
of the order $1/L$, and the exciton peak
is broadened and flattened because of the scattering
by chain ends.
If we choose the field-theory parameters
$\Delta$ and $A$ (the energy and conductivity scales)
to fit the finite-system DDMRG conductivity,
differences almost completely vanish for
$\Delta=0.669t$ and $A=1.02 e^2t$, as seen
in the inset of Fig.~\ref{fig:DMRGvsFT}.

As mentioned in Sec.~\ref{Subsec:intermediategap}
we have never observed more than one exciton below
the optical absorption continuum in the regime of intermediate
Mott gaps, down to $E_{\rm c}=0.4t$. Therefore,
we cannot evaluate the field-theory
predictions for the additional excitons and
the exciton-exciton continuum in the regime 
$\beta^2 \leq \frac{1}{3}$.
It remains conceivable, however, that smaller values
of $\beta^2$ can be reached in the model~(\ref{Hamiltonian})
in the limit $E_{\rm c}=\Delta \ll t$.  

\section{Exciton properties}
\label{Sec:excitonproperties}

In Wannier theory~\cite{Koch} exciton properties, such as size,
binding energy, effective mass, or optical weight,
are related by simple equations and exhibit a monotonic
behavior as a function of the Coulomb repulsion strength.
This simplicity is due to a drastic assumption of this theory:
optical excitations are represented by an electron 
(in the conduction band) and a hole (in the valence band), which are
completely independent from the system's other degrees of freedom.
The interaction with these degrees of freedom
is taken into account only through renormalized parameters
such as effective masses for the electron and hole, and 
an effective background dielectric constant for the Coulomb
interaction between electron and hole.

In a Mott insulator the exciton properties show a more
complex behavior. 
Although a Mott-Hubbard exciton can also be described as
a bound pair of excitations with opposite charges,
the Coulomb interaction at the same time determines 
the size of the Mott gap, the exciton properties, and the coupling 
of the exciton to the other electrons in the system. 
Therefore, an increase of the Coulomb interaction strength does
not simply bind the exciton more tightly, but also renormalizes
the gap and couples the particle-hole excitation more
strongly to the other electrons.
This leads to a non-monotonic behavior of Mott-Hubbard exciton
properties as a function of the Coulomb repulsion strength,
and even to an instability toward the formation
of excitonic strings or charge-density-wave droplets~\cite{EricUV}.

In the following two subsections we shall discuss the binding energy
and size of a Mott-Hubbard exciton using
analytical results in the limits of large and small Mott gaps,
and numerical results in the intermediate regime.

\subsection{Binding energy}
\label{Subsec:bindingenergy}

The exciton binding energy is usually defined as the energy difference 
between the exciton $\omega_{\rm ex}$ and the band edge of 
the particle-hole continuum $E_{\rm c}$
\begin{equation}
\delta E \, = \, E_{\rm c} - \omega_{\rm ex} \, .
\label{bindingenergy}
\end{equation}
In Wannier theory this quantity is the
energy required to break an exciton into
independent hole and electron.

\begin{figure}[ht]
\begin{center}
\epsfig{file=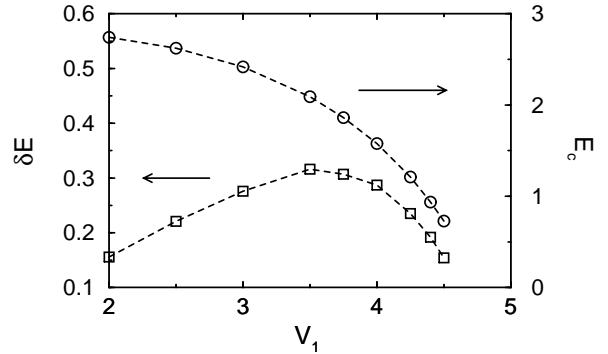,width=8cm}
\end{center}
\caption{Charge gap $E_{\rm c}$ (circles, right axis)
 and exciton binding energy
$\delta E$ (squares, left axis)
as a function of $V_1$ for $U=6t$ and $V_2=2t$. 
}
\label{fig:bindingenergy}
\end{figure}

In a Mott insulator this is not always warranted
as $\delta E$ does not necessarily 
correspond to the minimal energy required to break an exciton. 
For instance, as seen in~(\ref{2ndexc}),
there is an exciton {\em above} the band edge in the
strong-coupling limit $U \gg 4t > V_1$ ($V_2=0$).
Except for this strong-coupling limit, however, we
have only found excitons with an energy $\omega_{\rm ex}$
lower than the absorption band edge $E_{\rm c}$ in our study    
of the model~(\ref{Hamiltonian}).
Therefore, we assume that $\delta E$ is indeed the binding energy
of the single Mott-Hubbard exciton observed in the spectrum
of the Mott-insulating phase of~(\ref{Hamiltonian}). 

In the limit of a large Mott gap ($U \gg t, V_1, V_2$)
$\delta E = V_1 + 4t^2/V_1 - 4t$ for $V_1 \geq 2t$ and $V_2=0$,
see~(\ref{1stexc}).
The binding energy increases monotonically with $V_1$ 
but, clearly, it is not a good measure for the strength of
the Coulomb interaction as it is independent of~$U$ and
vanishes even when $V_1$~is still quite strong ($V_1 = 2t$).
For intermediate Mott gaps ($0.1 \leq E_{\rm c}/4t \leq 10$)
we have found that the binding energies (ranging from
$0.03t$ to $12t$) do not depend in a simple way on the 
Coulomb interaction strength. 
In particular, $\delta E$ has a non-monotonic behavior
as illustrated in Fig.~\ref{fig:bindingenergy}.
The binding energy $\delta E$ first increases with $V_1$ as 
long as the gap $E_{\rm c}$
remains essentially unchanged, then $\delta E$ and $E_{\rm c}$
decrease rapidly as $V_1$ approaches the critical surface
$U-2V_1+2V_2 \approx 0$.
It is likely that the same behavior also holds for small Mott gaps 
($E_{\rm c} = \Delta \ll t$), where field theory gives
$\delta E =  \Delta (1 - \sin(\pi \xi/2))$, $\xi \leq 1$,
for the first exciton, see~(\ref{exgap}).
If we take the weak-coupling result~(\ref{betasmallcouplings}) as an
indication of the qualitative dependence of $\beta$ on $V_1$, then
an increase of $V_1$ leads to smaller $\xi$,~(\ref{xi}), and thus to a
larger binding energy. 
On the other hand, an increase of $V_1$ can also leads to a sharp
{\sl decrease\/} of the gap $\Delta$ close to the critical surface
$U-2V_1+2V_2 \approx 0$, if we again take the weak-coupling 
results~(\ref{FTgap1}) and~(\ref{FTgap2}) to be indicative of the 
qualitative
dependence of $\Delta$ on $V_1$.
Depending on the other parameter values, one of these two
effects dominates and leads to an increase or decrease
of the binding energy when $V_1$ becomes larger.

In summary, our analysis shows that the binding energy of a 
Mott-Hubbard
exciton does not provide a good estimate of the
Coulomb interaction strength in a Mott insulator.
Of course, a large gap $E_{\rm c}$ or binding energy $\delta E$
requires a strong Coulomb interaction. In general, however, a small
$E_{\rm c}$ or $\delta E$ are {\sl not\/} an
indication for a weak electron-electron
interaction. In contrast to the views of Ref.~\onlinecite{HeegerSari},
even a ``small'' exciton binding energy,
of the order of~$0.1\, {\rm eV}$ in some
$\pi$-conjugated systems, does {\sl not\/} imply that electron-electron
interactions are small in these materials.

\subsection{Size}
\label{Subsec:excitonsize}

In the limit of large Mott gaps ($U \gg t, V_1, V_2$)
an optical excitation is
simply made of a hole and a double occupancy in a background
of singly occupied sites (Sec.~\ref{Subsec:largegap}).
The probability of finding the hole and double occupancy
at a distance $x$ in an optical excitation
is given by the correlation function
\begin{equation}
C_{\rm hd}(x) \, = \, \langle \hat{n}_{l, \uparrow}
\hat{n}_{l, \downarrow} (1-\hat{n}_{l+x, \uparrow}) 
(1-\hat{n}_{l+x, \downarrow}) \rangle \, ,
\label{hdcorrelation}
\end{equation}
where $\langle \dots \rangle$ means the expectation value
in the corresponding excited $N$-electron eigenstate.
The average distance between hole and double occupancy
is
\begin{equation}
\zeta_{\rm hd} \, = \, \frac{\sum_x C_{\rm hd} (x)  |x|}
{\sum_x C_{\rm hd} (x)}
\, \, .
\label{excitonsize}
\end{equation}
If the hole and double occupancy are not bound, 
this quantity diverges as the system size $L$ goes to infinity.
For an exciton, $\zeta_{\rm hd}$ remains finite as $L\rightarrow \infty$
and can be interpreted as the exciton size.
For $V_1 \geq 2t$ and $V_2=0$ an analytical calculation gives the 
exact result
\begin{equation}
C_{\rm hd}(x \neq 0) \propto \exp(-\kappa|x|)
\end{equation} 
with
$\kappa = 2\ln(V_1/2t)$ for the lowest exciton. The exciton size is
then 
\begin{mathletters}
\label{exactexcitonsize}
\begin{eqnarray}
\zeta_{\rm hd} &=& \frac{1}{1-e^{-\kappa}} \label{exactexcitonsizeA}\\
  &=& \frac{1}{1-(2t/V_1)^2} \ .
  \label{exactexcitonsizeB}
\end{eqnarray}\end{mathletters}%
$\zeta_{\rm hd}$ diverges as
$V_1$ approaches the critical value ($V_{\rm c} = 2t$),
below which the pair is not bound,
and tends to unity for strong coupling, $V_1/2t \rightarrow \infty$.
Using a density-density correlation function~\cite{mazumdar,bursill}
yields equivalent results.

Unfortunately, the correlation function for hole and double occupancy
and the density-density correlation 
functions yield unclear results 
in the regime of intermediate Mott gaps because
the ground state already contains a finite density of 
holes and double occupancies when $E_{\rm c}/t$ is finite.
The function~(\ref{hdcorrelation}), for instance, tends
to a finite value $d^2$ as $x \to \infty$, which is
determined by the density of doubly 
occupied sites $d = \langle \hat{n}_{l, \uparrow}
\hat{n}_{l, \downarrow} \rangle$.
These quantum charge fluctuations hide 
the weak exciton contribution to~(\ref{hdcorrelation}). 

\begin{figure}[ht]
\begin{center}
\epsfig{file=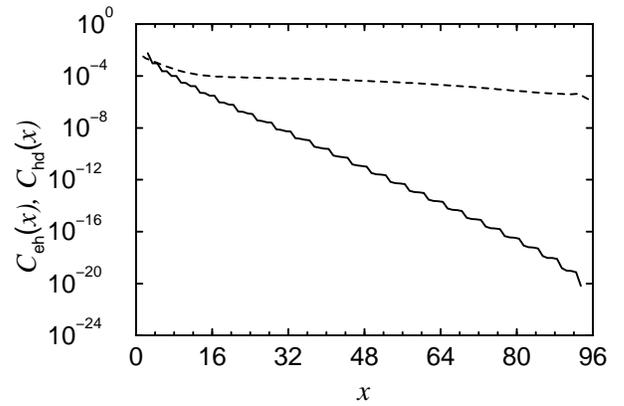,width=8cm}
\end{center}
\caption{Correlation functions for a hole and double occupancy,
 $C_{\rm hd}(x)$ (dashed), and for an 
electron-hole excitation, $C_{\rm eh}(x)$ (solid),
calculated for the lowest exciton in a 96-site system with 
$U=40t$, $V_1=2.5t$, and $V_2=0$. 
The center of the pair is in the middle of the lattice.}
\label{fig:excitoncorrelation1}
\end{figure}

In Fig.~\ref{fig:excitoncorrelation1} we show 
the correlation function $C_{\rm hd}(x)$
calculated with DMRG
for the lowest optically excited state 
in the model~(\ref{Hamiltonian})
with $U=40t$, $V_1=2.5t$, and $V_2=0$. 
This system is in the regime 
of large Mott gaps with $E_{\rm c}\approx 36.14t$,
and is well described by our strong-coupling 
analysis, which predicts an exciton with a size 
$\zeta_{\rm hd} \approx 2.78$
as the lowest optically excited state.
However, one clearly sees in Fig.~\ref{fig:excitoncorrelation1}
that $C_{\rm hd}(x)$ remains finite 
for large $x$. This wrongly suggests that the 
optically generated hole and double occupancy 
are not bound in this excited state.
Similar problems arise with a density-density correlation function.
Taking the difference between the correlation 
functions for an excited state 
and for the ground state, as in Ref.~\onlinecite{bursill}, 
does not provide better results.

A suitable quantity for our analysis 
is the correlation function for electron-hole
excitations 
\begin{equation}
C_{\rm eh}(x) = \left | \left \langle n \left | \hat{P}_{l,l+x}
+ (-1)^{|x|} \hat{P}_{l+x,l}
 \right| 0 \right \rangle \right |^2  \; ,
\label{exccorrelation}
\end{equation}
where $|0\rangle$ is the ground state, $|n\rangle$ is the
excited state under investigation, and 
\begin{equation}
\hat{P}_{i,j}  = 
\sum_{\sigma}  \hat{c}^{+}_{i, \sigma} \hat{c}_{j, \sigma} 
\label{correlationoperators}
\end{equation}
creates an electron at site~$i$ and a hole at site~$j$.
Obviously, $C_{\rm eh}(x)$ evaluates the importance
of an electron-hole excitation with distance~$x$
in the excited state $|n\rangle$.
This approach has already been
used to study the structure of excited states in 
semiempirical calculations
for ladder-type PPP oligomers~\cite{joerg}.
Here, we have calculated this correlation function using
the symmetrized DMRG method
to analyze the structure  of excited states in the lattice 
model~(\ref{Hamiltonian}).

An average electron-hole distance $\zeta_{\rm eh}$
can be defined  by substituting $C_{\rm eh}(x)$
for $C_{\rm hd}(x)$ in Eq.~(\ref{excitonsize}).
We have found that this method predicts
correctly whether a hole-double occupancy pair
is bound or unbound 
in the limit of large Mott gaps ($U \gg t,V_1,V_2$).
The advantage of the correlation function $C_{\rm eh}(x)$
over $C_{\rm hd}(x)$ (or a density-density correlation function)
becomes obvious in a system with a {\sl finite\/} Mott gap.
In Fig.~\ref{fig:excitoncorrelation1} one sees that
$C_{\rm eh}(x)$ decreases exponentially and thus allows us to
identify a bound excitation, while  
the correlation function 
between hole and double occupancy $C_{\rm hd}(x)$
is dominated by the finite ground-state
density of holes and double occupancies. 
Therefore, we think that 
the correlation function
for electron-hole excitations $C_{\rm eh}(x)$ 
and the analysis of the average electron-hole 
distance $\zeta_{\rm eh}$ 
provide a reliable approach 
to distinguish an exciton from an unbound particle-hole
excitation in correlated systems.
In any case, this approach is more reliable than
schemes based on the correlation function for hole and double 
occupancy or a density-density correlation function.

However, this approach cannot be used to determine
the size of a bound pair accurately because $C_{\rm eh}(x)$
is affected by very strong 
short-range fluctuations due to spin correlations
and lattice and chain-end effects.
[In our figures we show a running average of $C_{\rm eh}(x)$.]
Instead, 
we evaluate the exciton size $\zeta_{\rm ex}$ as the average
electron-hole distance [as in Eq.~(\ref{excitonsize})]
calculated for the exponential
function which gives the best fit to $C_{\rm eh}(x)$
\begin{mathletters}
\label{howtodefineexcitonsize}
\begin{eqnarray}
C_{\rm eh}(x) \propto
 \exp(-\lambda|x|)\; , 
\label{howtodefineexcitonsizeA}\\
\zeta_{\rm ex} =   \frac{1}{1-\exp(-\lambda)}
\; .
\label{howtodefineexcitonsizeB}
\end{eqnarray}\end{mathletters}%
In the limit of large Mott gaps we have found 
that this approach reproduces the exact result~(\ref{exactexcitonsizeB}) 
for the exciton size,
i.e., $\zeta_{\rm ex} \approx \zeta_{\rm hd}$.
Furthermore, applying this analysis to systems with large but finite
gaps yields results in agreement with our strong-coupling
analysis. For instance, using the data for
$U=40t$ and $V_1=2.5t$ in 
Fig.~\ref{fig:excitoncorrelation1}
our analysis gives $\zeta_{\rm ex}= 2.99$. 
This value agrees within 10\% with
the analytical result~(\ref{exactexcitonsizeB})
for $U/t\to\infty$ and $V_1=2.5t$, 
$\zeta_{\rm hd}= 2.78$.
The difference can be explained as a correction of the order
$4t/U$.
Therefore, we think that the analysis~(\ref{howtodefineexcitonsize})
of the correlation function for electron-hole excitations
allows one to determine reliably the size of an exciton
in a correlated system.

\begin{figure}[ht]
\begin{center}
\epsfig{file=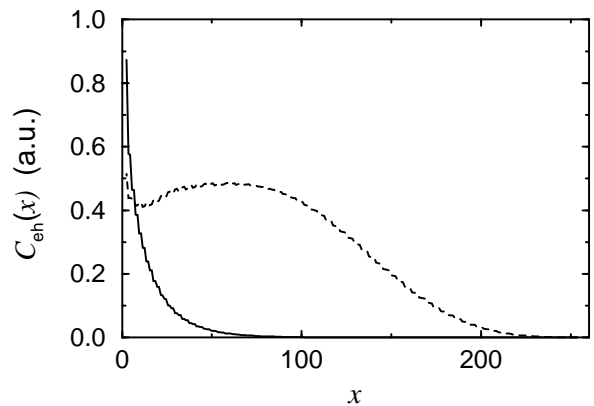,width=8cm}
\end{center}
\caption{Correlation function 
$C_{\rm eh}(x)$ of the first optically
excited state on a 256-site lattice
for two typical cases:
an exciton (solid) for $U=4t$, $V_1=2.75t$, and $V_2=1.5t$, 
and an unbound particle-hole pair (dashed)
for $U=3.5t$, $V_1=1.4t$, and $V_2=0$.
The center of the pair is in the middle of the lattice.
}
\label{fig:excitoncorrelation2}
\end{figure}

We have calculated the average electron-hole
distance $\zeta_{\rm eh}$
of the lowest optically excited states 
for various values of the 
parameters $U$, $V_1$, and $V_2$
in the regime of intermediate Mott gaps.
The analysis of $\zeta_{\rm eh}$ yields
predictions about the presence of bound particle-hole pairs 
which always agree with 
the predictions based on the analysis of the
excited-state energies. 
For an excited state above the absorption band threshold we have found
that $\zeta_{\rm eh}$ diverges with system size, confirming
that this excitation is made of an unbound particle-hole pair.
A representative example for the various possible shapes
of $C_{\rm eh}(x)$ is shown in 
Fig.~\ref{fig:excitoncorrelation2}.
For an unbound pair one clearly sees 
that $C_{\rm eh}(x)$ remains finite for very 
large~$x$ of the order of the system size~$L$.
$C_{\rm eh}(x)$ only goes to zero as $x$ approaches the system size
($L=256$ in this example) because of chain-end effects.

In contrast, when an excited
state lies below the charge gap $E_{\rm c}$,
we have found that $\zeta_{\rm eh}$ remains finite for an infinite
system, confirming that this excited state is made of a bound
particle-hole pair (exciton). In this case, $C_{\rm eh}(x)$ 
decreases exponentially and we can use the 
analysis~(\ref{howtodefineexcitonsize})
to determine the exciton size $\zeta_{\rm ex}$. 
A typical example is also shown
in Fig.~\ref{fig:excitoncorrelation2}.
We note that both examples in Fig.~\ref{fig:excitoncorrelation2} 
correspond to systems with a gap $E_{\rm c} \approx 0.66t$,
for which the field theory of Sec.~\ref{Sec:FT}
is still valid and supports our identification
of bound and unbound excitations. 

The exciton sizes $\zeta_{\rm ex}$ observed range from slightly more
than one lattice spacing $a_0$ to about $20 a_0$.
As the binding energy $\delta E$, the exciton size
$\zeta_{\rm ex}$ does not depend in a simple
way on the Coulomb interaction strength
and even displays a non-monotonic
behavior as a function of $V_1$, first decreasing
as $V_1$ increases, then sharply increasing
as $V_1$ approaches the critical regime
$U-2V_1+2V_2 \approx 0$.
As expected, $\zeta_{\rm ex}$ becomes larger and seems to
diverge as $\delta E$ vanishes, while it diminishes
when $\delta E$ becomes larger
if the gap $E_{\rm c}$ is kept (approximately) constant.

We note that the operator used in~(\ref{exccorrelation}) 
is antisymmetric
with respect to charge conjugation, as is the current 
operator~(\ref{currentop}).
Thus, $C_{\rm eh}(x)$ is non-zero only for those excited states
$|n\rangle$ (including the excited states contributing
to the linear optical conductivity) whose parity
$P_{\rm c}$ under charge conjugation is opposite to that of the 
ground state.
When a minus sign is substituted for the $+$ sign 
in~(\ref{exccorrelation}), $C_{\rm eh}(x)$ is non-zero for 
excited states which have the same parity under charge 
conjugation as the ground state.
One can thus extend this scheme 
to excited states which contribute to the non-linear
optical properties of a system. 

In the field-theory limit, a measure of the size of an exciton can be
obtained in the following way. For asymptotically large separation
between particles, the holon-antiholon wave function has
the form
\begin{eqnarray}
\Psi_{x_1\ll x_2}(x_1,x_2)&=&\exp(i p_1x_1+ip_2x_2)\nonumber\\
&&+S_{\rm R}(p_1,p_2)\exp(i p_2x_1+ip_1x_2)\; ,\nonumber\\[6pt]
\Psi_{x_1\gg x_2}(x_1,x_2)&=&S_{\rm T}(p_1,p_2)\exp(i p_1x_1+ip_2x_2)\; ,
\end{eqnarray}
where $S_{\rm R,T}$ are two-particle $S$-matrix elements corresponding to
reflection and transmission respectively.
In terms of rapidity variables
defined by $p_j=M\sinh\theta_j/v_{\rm c}$, the $S$-matrix only depends 
on the
difference $\theta_{12}=\theta_1-\theta_2$ and has poles at
\begin{equation}
\theta_{12}=i\pi(1-n\xi)\ .
\end{equation}
An exciton state with energy $M_n\cosh\theta$ and momentum
$M_n\sinh\theta/v_{\rm c}$ is obtained by choosing
\begin{eqnarray}
p_1&=&\frac{M}{v_{\rm c}}\sinh(\theta+i\frac{\pi-\pi n\xi}{2})\ ,
\nonumber\\[6pt]
p_2&=&\frac{M}{v_{\rm c}}\sinh(\theta-i\frac{\pi-\pi n\xi}{2})\ .
\end{eqnarray}
Inserting these values in the wave function yields an exponential
decay in $|x_2-x_1|$ with correlation length
\begin{mathletters}
\label{excitonradiusFT}
\begin{eqnarray}
\zeta^{\rm FT}_{n}&=&\frac{v_{\rm c}}{M\cos \pi n\xi/2}\; ,
\label{excitonradiusFTa} \\
&=& \frac{2v_{\rm c}}{\sqrt{\Delta^2-M_n^2}} \; .
\label{excitonradiusFTb}
\end{eqnarray}\end{mathletters}%
We see that this size diverges when we approach from below the
coupling constant $\xi$, at which the $n$-th exciton is first
formed. For example, the first exciton splits off the holon-antiholon
continuum at $\xi=1$ and its size displays a square-root
divergence for $\xi\to 1$

\begin{equation}
\zeta^{\rm FT}_{1}\propto \sqrt{\frac{1}{\Delta-M_1}}
\quad ; \quad M_1\to \Delta^- \; .
\end{equation}

The non-monotonic behavior of the exciton size as a function
of the Coulomb interaction strength is also observed
in the small-gap regime. Equation~(\ref{excitonradiusFTa}) 
shows that $\zeta_1^{\rm FT}$ will decrease if the 
diminution of $\xi$ dominates 
when the lattice parameter
$V_1$ increases,
but becomes greater if 
the increase of the velocity $v_{\rm c}$ and the
reduction of the gap $\Delta$ prevail
(see Sec.~\ref{Sec:FT}).

To compare the field-theoretical predictions with our numerical
method based on the correlation function
$C_{\rm eh}(x)$ we have numerically calculated the
exciton energy $\omega_{\rm ex}$ and 
size $\zeta_{\rm ex}$
for several values of the parameters $U$, $V_1$, and $V_2$,
corresponding to gaps $\Delta=E_{\rm c}$ of the order
of $0.6t$.  
The exciton sizes range from $10 a_0$ to $20 a_0$.
We have found that in all cases our numerical results fulfill
the field-theory relation~(\ref{excitonradiusFTb})
within 15\% if we use $v_{\rm c} = 2t a_0$ as the charge velocity.
This good agreement shows that our 
definitions~(\ref{howtodefineexcitonsize}) 
and~(\ref{excitonradiusFT}) are mutually consistent
in the regime of small Mott gaps. Moreover, 
the numerical method yields reasonable estimates of the exciton size
even in the regime where such calculations become laborious.

\section{Conclusions}
\label{Sec:conclusions}

We have investigated excitons in the optical conductivity
spectrum of one-dimensional Mott insulators using
two new reliable methods, the dynamical density-matrix
renormalization group and the field-theoretical
form-factor bootstrap approach, supplemented by two
established techniques, a strong-coupling analysis
and the symmetrized DMRG.  
Mott-Hubbard
excitons can be understood with the simple picture 
of a bound pair of spinless bosonic excitations with opposite charge,
in analogy to the bound electron-hole pair of
Wannier exciton theory.
However, the properties of Mott-Hubbard excitons are not as simple as
those of Wannier excitons because
both the Mott gap and the force between charged excitations
originate from the Coulomb repulsion between electrons
and are thus interdependent. 
In particular, the smallness of an exciton binding
energy is no indication for the strength
of the Coulomb repulsion in a Mott insulator. 

We may compare our results to experiments for polydiacetylene
chains in their monomer matrix. For an estimate of the exciton
size we use the field-theoretical result~(\ref{excitonradiusFT})
which is fairly independent of the details of the lattice
structure and the range of the electron-electron interaction.
We put $v_{\rm c}\approx v_{\rm F}= 2ta_0$ 
with a $\pi$-electron band width 
$W=4t\approx 10\, {\rm eV}$~\cite{Salaneck},
$\Delta=\omega_{\rm ph}=2.4\, {\rm eV}$,
and $M_1=\omega_{\rm ex}=1.9\, {\rm eV}$ from experiment
for 3BCMU-PDA~\cite{Weiser2}. The field-theoretical prediction
for the exciton size,
\begin{equation}
\zeta_1^{\rm FT} = a_0 \frac{W}{\sqrt{\omega_{\rm ph}^2-
\omega_{\rm ex}^2}}
\approx 7 a_0 \approx 9\, {\rm \AA}
\label{data}
\end{equation}
is in fair agreement with the experimental value 12~\AA\
if we put $a_0=1.3$~\AA\ for the average bond length between the 
carbon atoms on the PDA chain.
The difference can be attributed to the limitations of 
this simple calculation 
or the uncertainty of the order of  20\%
for the experimental values\cite{Weisercommunication}.
We therefore conclude that our many-body approach 
can be applied successfully to real polymers.

Although the particular model studied here is too simple to
describe conjugated polymers accurately, we believe that the 
concept of Mott-Hubbard excitons is relevant
for these materials as the electron-electron
interaction significantly contributes
to the formation of their optical gap.  
In any case, this approach is more realistic
than the oversimplified Wannier theory 
and other simple approaches that neglect or minimize
the role of electronic correlations in conjugated polymers. 
The many-body methods used in this work can be 
applied (and in part, have already been applied)
to more realistic models taking into account
the polymer geometrical structure and the electron-phonon
interaction, and possibly additional perturbations
such as interchain couplings. 
We think
that the optical properties of conjugated polymers
will be successfully investigated using a combination
of these many-body methods.

\acknowledgments

We gratefully acknowledge helpful discussions with R.~Bursill 
and G.~Weiser.

\end{multicols}

\end{document}